\documentclass[12pt,preprint]{aastex}
\usepackage{natbib}
\usepackage{graphicx,graphics,subfigure}
\bibpunct{(}{)}{;}{a}{}{,}

\slugcomment{To appear in the Astrophysical Journal}

\shorttitle{A bent DLRS in the A1763 filament}
\shortauthors{Edwards et al.}

\begin{document}

\title{The first bent double lobe radio source in a known cluster filament: Constraints on the intra-filament medium}

\author{Louise O. V. Edwards and Dario Fadda
\affil{NASA Herschel Science Center, Caltech 100-22, Pasadena, CA 91125}}
\email{louise@ipac.caltech.edu}
\author{David T. Frayer
\affil{NRAO, Green Bank, WV, 24944}}

\begin{abstract}

We announce the first discovery of a bent double lobe radio source (DLRS) in a known cluster filament. The bent DLRS is found at a distance of 3.4$\,$Mpc from the center of the rich galaxy cluster, Abell~1763. We derive a bend angle $\alpha$=25$^{o}$, and infer that the source is most likely seen at a viewing angle of $\Phi$=10$^{o}$. From measuring the flux in the jet between the core and further lobe and assuming a spectral index of 1, we calculate the minimum pressure in the jet, (8.0$\pm$3.2)$\times$10$^{-13}$ dyn/cm$^{2}$, and derive constraints on the intra-filament medium (IFM) assuming the bend of the jet is due to ram pressure. We constrain the IFM to be between (1-20)$\times$10$^{-29}$ gm/cm$^{3}$. This is consistent with recent direct probes of the IFM and theoretical models. These observations justify future searches for bent double lobe radio sources located several Mpc from cluster cores, as they may be good markers of super cluster filaments.

\end{abstract}

\keywords{galaxies: clusters: intracluster medium --- galaxies: clusters: individual (Abell 1763, Abell 1770) --- large-scale structure of universe --- radio continuum: galaxies}

\section{Introduction}

Bent double lobe radio sources (DLRSs) are rare, found in only 6\% of clusters \citep{bla01}. The bend is thought to be caused by ram pressure of the host cluster or group medium \citep{owe76}, by an interaction with the host galaxy, or by merger activity with another galaxy \citep{rec95}. The environment surrounding a bent DLRS has been considered for decades \citep{kun80,gow84,bla00,dev06}. However, this only includes a region out to the virial radius of the system, most often an intermediate density group \citep{cro08,zir97}. Cluster scale filaments are also postulated to be host to bent DLRSs \citep{fre08}, but the  classification of such structures is observationally difficult as the galaxy density is low and the physical scale on the sky is vast. 

We have been studying the super structure comprised of the rich cluster Abell~1763 and the poor cluster Abell~1770. A filament which connects the two was discovered using mid-IR observations from the {\it Spitzer} Space Telescope \citep{fad08}, confirmed by collecting hundreds of member galaxy redshifts, and observed in 13 optical and IR bands \citep{edw10}. With new VLA data \citep{edw10b} we have discovered a DLRS (F1) located in the cluster feeding filament, well beyond the virial radius.

We announce this first detection of a bent DLRS in a known cluster filament. We use the properties of the bend to constrain the density of the surrounding intra-filament medium (IFM). We adopt a cosmology with $\Omega$$_{m}$=0.3, $\Omega$$_{\Lambda}$=0.7 and H$_{o}$=70$\,$km/s/Mpc where one arcsec corresponds to 3.7$\,$kpc at a redshift of 0.23.

\section{Observations and Data Reduction}

\begin{figure}
   \center
  \epsscale{1.0}
     \plotone{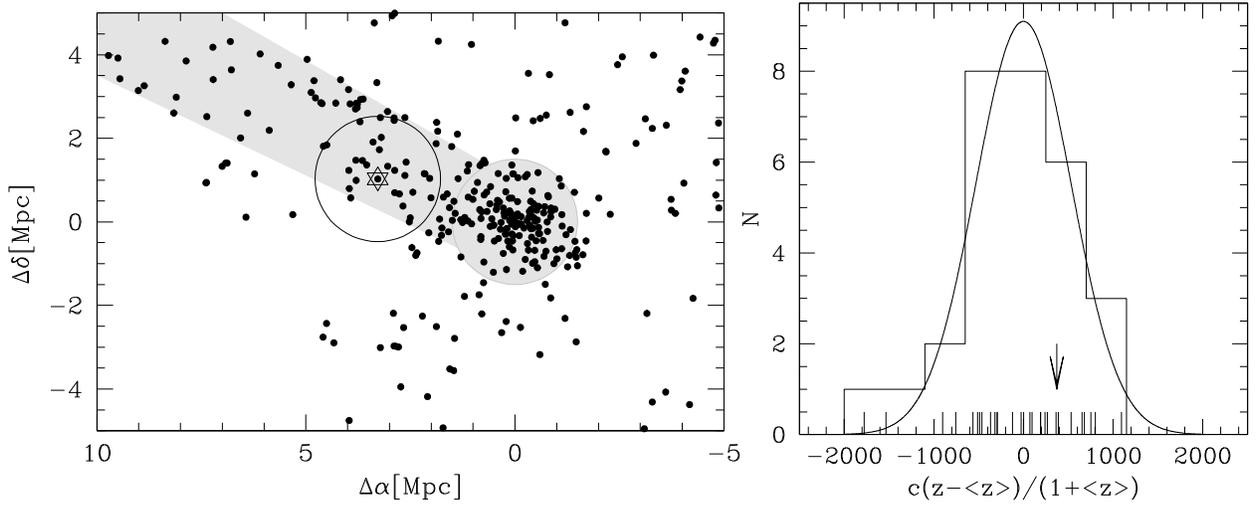}
    \caption[]{{\bf The velocity of F1 with respect to the filament} Left (a): The spectroscopically confirmed members are shown as black dots, with F1 highlighted as a star. The location of the filament is shaded in grey. The large grey circle marks the cluster R$_{500}$. The large black circle has a radius of 1.5$\,$Mpc and inscribes the galaxies included in the histogram on the right. Right (b): A histogram of the velocities of the 29 galaxies relative to their average. It includes galaxies within 1.5Mpc of F1 and with redshifts between 0.025-0.24. The Gaussian corresponds to the normalized distribution of velocities ($\sigma_{l.o.s} = 532\,$km/s). The vertical lines at the bottom of the histogram mark the relative velocity of each galaxy, F1 is highlighted with an arrow. 

\label{posvel}}
\end{figure}

The data reduction for the VLA 1.4GHz maps is discussed in detail in the companion paper, Edwards et al. (2010b). Both the FIRST survey (rms = 120$\mu$Jy/Beam) and NVSS (rms = 450$\mu$Jy/Beam) detect this source as an unresolved point at 1.4GHz. We achieve a spatial resolution of 5$\,$arcsec, and in our deeper observations (rms = 28$\mu$Jy/Beam) two diffuse lobes on either side of the point source appear.

\begin{figure}
   \center
  \epsscale{1.0}
     \plotone{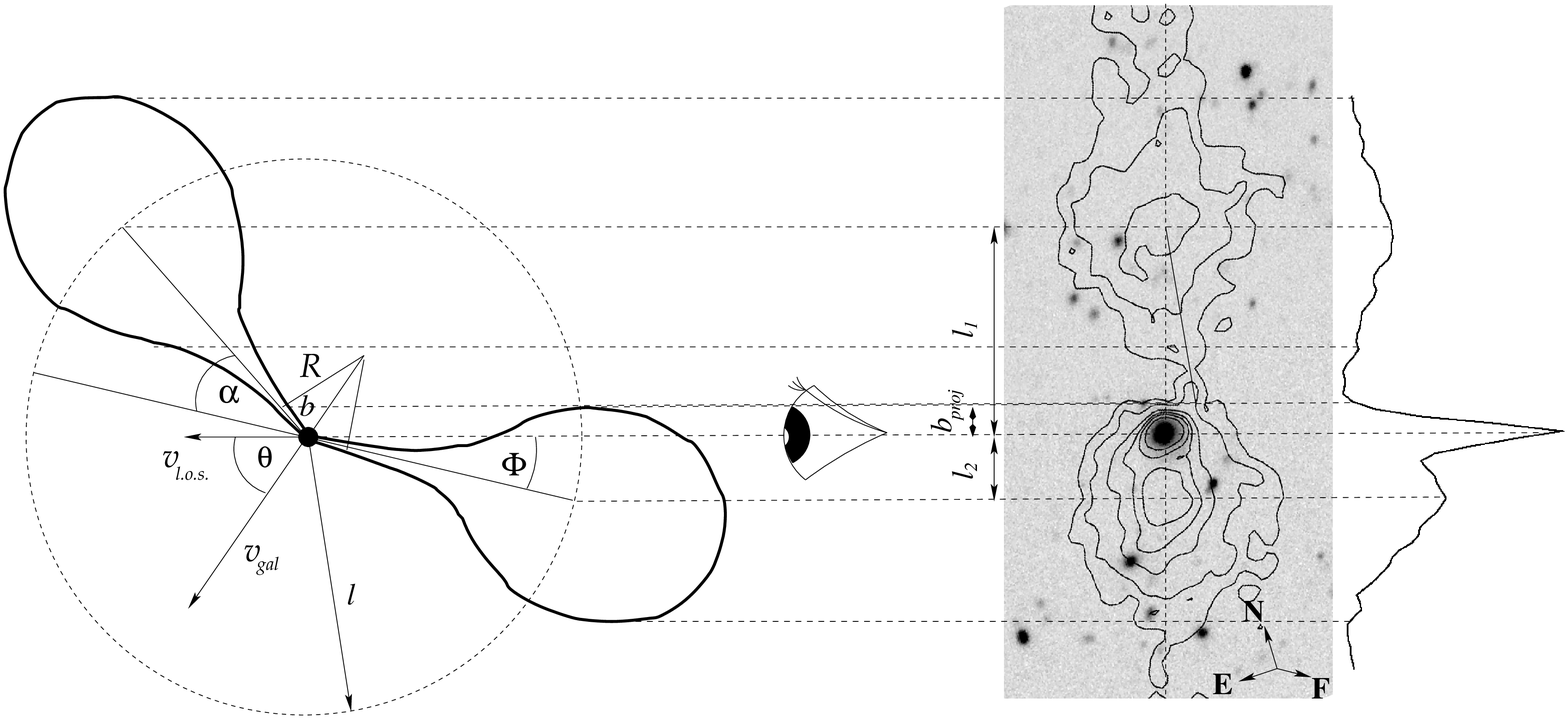}
    \caption[]{{\bf Geometry of the bent DLRS}  The R-band image surrounding the radio-excess galaxy, F1, is shown in negative grey-scale.  The black contours show the large lobes of radio emission, the beam is 5$\,$arcsec. The projected core-lobe distances are $l_{1}$ and $l_{2}$. A cut through the major axis of the radio source is shown at the far right of the image. The large inscribed circle which defines the jet length, $l$, sets the various angles and projected lengths.  The bend angle, $\alpha$, bend length, $b$, its projection, $b_{proj}$, and the orientation angle, $\Phi$ are as shown. We also show the curvature radius, R, as described in the text. From the galaxy redshift we can determine the line-of-sight velocity, $v_{l.o.s.}$, which is the line-of-sight component of the true galaxy velocity ($v_{gal}$cos($\theta$)). To highlight how straight the upper jet is, we trace a thin line which joins the upper lobe to the beginning of the bend.
\label{bragn}}
\end{figure}

\section{Results}
\subsection{Physical properties of the source}

The physical location of the DLRS, named F1 in \citet{edw10b}, is at 13h36m36.6s +41d04m34.7s (J2000). Figure~\ref{posvel}a shows the relative distance of F1, $\sim$15.3$^{\prime}$ (3.4$\,$Mpc) from Abell~1763's brightest cluster galaxy, well outside the virial radius of either cluster and along the galaxy filament.  

F1 is a radio galaxy with a 1.4$\,$GHz power of 9.6$\times$10$^{23}$ Watts/Hz. The flux is dominated by an unresolved core and includes two diffuse lobes which have an intensity ratio of 2.25, clearly visible in Figure~\ref{bragn}. The lobe closest to the core is not only brighter, but also more compact than the elongated further lobe. We measure the core-lobe distance $l_{1}$=12.5$^{\prime\prime}$ (171$\,$kpc), and the lobe distance ratio to be 3.55. Based on the FRI/FRII power break \citep{led96}, and the fact that the source is core dominated \citep{fan74}, we classify F1 as an FRI.

As the two lobes are aligned, the source is probably nearly along the line-of-sight. However, as part of the jet does not follow the source axis, we include the caveat that our model may be a simplified version of a more complex reality.

The modest densities expected in the IFM allow for two physically motivated constraints. First, we assume that the two core-lobe lengths are intrinsically equal, which implies a bend in the arms unseen because of our viewing angle. This is supported by the jets being ejected far from the host galaxy, and along the minor axis, implying also a lack of interaction with the host. Second, the bend angle, $\alpha$, is unlikely so extreme as to be greater than 90$^{o}$. Figure~\ref{posvel}b shows a histogram of velocities for galaxies within a 1.5$\,$Mpc from F1. The positive relative velocity means that the galaxy is headed away from the observer. Taken in combination with the fact that the far lobe is more extended with a fainter peak flux, we take the further lobe to be oriented with its jet away from the observer, and the nearer lobe with its jet oriented towards the observer (as in Figure~\ref{bragn}).

\subsubsection{Deducing the intrinsic bend}

The bottom of Figure~\ref{bendang} shows our model for the bend angle $\alpha$=90$^{o}$-arcsin($l_{2}$/$l$)-arccos($l_{1}$/$l$), where $l$ is the true core-lobe distance. We take the minimum core-lobe distance to be 171$\,$kpc as measured, and calculate the bend angle as a function of intrinsic jet length. Typical bent DLRSs have jet lengths of under $\sim$ 400 kpc \citep{gow84}, and a reasonable minimum core-lobe length is $\sim$250$\,$kpc, so we take $\alpha = $20-30$\sim$25$^{o}$. This is consistent with models of \citet{lis94} based on a sample of double lobe radio sources in the sky which suggest intrinsic $\alpha$ between 0-25$^{o}$. 

\subsubsection{Orientation of the bent  DLRS}

We calculate the orientation angle of the approaching jet, $\Phi$, as a function of jet length, $\Phi$=arcsin($l_{2}$/$l$). Taking the same range for jet lengths as before, $\Phi$$\sim$7-12$^{o}$ as shown in the middle panel of Figure~\ref{orient}. This is also consistent with the \citep{lis94} models that derive best fit viewing angles of greater than 50$^{o}$ from the line-of-sight to the receding jet for their radio galaxies (180$^{o}$-$\alpha$-$\Phi$=145$^{o}$, in our case).

\subsubsection{Jet curvature}

\begin{figure}
   \center
  \epsscale{1.0}
     \plotone{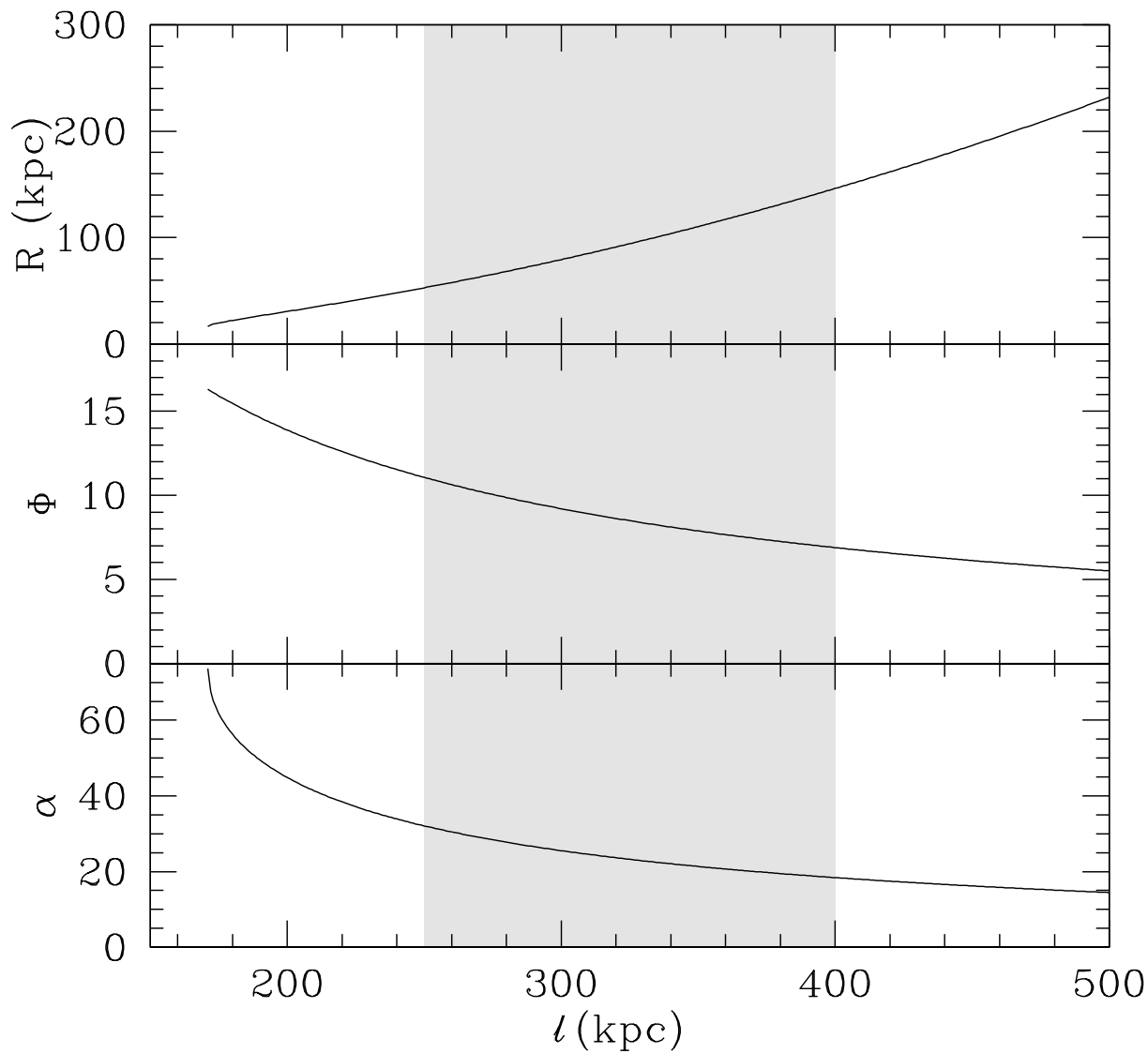}
    \caption[]{{\bf Geometrical Models} Top:  The radius of curvature, R, as a function of jet length ($l$). The curve delineates the upper bound of acceptable R, based on the upper limit of the projected bend length, $b_{proj}$. Middle: Leading jet orientation ($\Phi$) as a function of jet length. Bottom: Bend angle ($\alpha$) of the jet as a function of $l$. All angles are defined as in Figure~\ref{bragn}. Acceptable values of $l$ include the shaded region with $l$=250-400$\,$kpc. This range of $l$ defines our constraints on R, $\Phi$, and $\alpha$.
\label{bendang}
\label{orient}
\label{radcurv}} 
\end{figure}

The radius of curvature, R, of the bent DLRS can be calculated as a function of bend angle (and therefore of $l$). Assuming the jet starts to curve at a length $b$ along the jet, R can be defined as in Figure~\ref{bendang}, R=$b$/[cos(180-$\alpha$/2)]. Because the jet is essentially straight out to the contours of the lower lobe, we have an observational constraint on the upper limit of the projected component, $b_{proj}$. Therefore the line in the top panel of Figure~\ref{orient} defines the maximum values of the curvature radius at different $l$. A minimal value is estimated from the literature as $\sim$15$\,$kpc (see for example, O'Donohue et al. 1993).

\subsection{A probe of the IFM}

The presence of a bent DLRS in and of itself suggests an interplay between the radio source and its surroundings. Ram pressure is likely the cause of the bent morphology, especially in luminous X-ray clusters \citep{bla01} where the ICM is dense. But it is not clear whether the much less dense IFM could similarly influence the jets.

Presently, we use the source geometry to derive constraints on the density of a posited IFM. In addition, we calculate a merger timescale for the closest companions which is much longer than the lifetime of the jet particles.

\subsubsection{Constraints on the jet properties and the intra-filament medium}

We use the Euler equation as in \citep{beg79} and \citep{fre08} to estimate $\rho_{IFM}$. 

\begin{equation}
\frac{\rho_{IFM}\mathnormal{v}^{2}}{h}=\frac{\omega \Gamma^{2} \beta^{2}}{R}, 
\end{equation}

This model assumes that ram pressure from the relative motion between the source and its surrounding medium causes the bend in the radio jets. We measure the width of the jet, $h$, as 48$\,$kpc, from our map. Although, we point out this is an upper bound as  our 20$\,$cm observations may not resolve the true width of the jet. The term ($\omega\Gamma^{2}$)($\beta^{2}$) includes the enthalpy, $\omega$, and depends on the pressure inside the jets and the jet velocity ($\omega$=4$\,$P$_{min}$). $\Gamma$ is the relativistic factor, (1-$\beta^{2}$)$^{-1/2}$, and $\beta=v/c$. We adopt the same value of $\beta= 0.54 \pm 0.18$ as in \citet{fre08} as it is consistent for speeds of FR~I sources with both straight \citep{ars04} and wide angle tail \citep{jet06} morphologies. R is the radius of curvature and not straightforward to measure, therefore, we calculate our results for several values. The pressure in the jets can be calculated as in \citet{pac70} and \citet{smo08}, and depends on the jet flux density and the spectral index, \begin{math}\mathnormal{a}\end{math}. 

\begin{equation}
\mathrm{P}_{min}=\frac{7}{9}\frac{\mathrm{B_{me}^{2}}}{8 \pi}, 
\end{equation}

where,

\begin{eqnarray}
\lefteqn{\mathrm{B_{me}^{2}  = 5.69 \times 10^{-5} [\frac{1+k}{\eta}(1+z)^{3+\mathnormal{a}}}   }\nonumber \\
& & \mathrm{\times \frac{1}{\Theta_{x}\Theta_{y} l sin^{3/2} \Phi}\frac{F_{o}}{\nu_{o}^{\mathnormal{a}}}\frac{\nu_{2}^{1/2 - \mathnormal{a}} - \nu_{1}^{1/2 - \mathnormal{a}}}{(1/2) - \mathnormal{a}}]}.
\end{eqnarray}

Radio observations for this source exist only at one frequency, so we can not measure the spectral index. It is not straightforward to assume a value as there can be wild variability, even within a single source. We therefore calculate the jet power and $\rho$$_{IFM}$ for several values of \begin{math}\mathnormal{a}\end{math}.

We measure a flux density of 0.75$\pm$0.26$\,$mJy/beam in the jet going towards the more distant lobe, $\Theta_{x}$=6.8$\,$kpc and $\Theta_{y}$=4.8$\,$kpc, and assume cylindrical symmetry in the jet. We calculate the bolometric radio luminosity between frequencies of $\nu_{o}$=0.01 and $\nu_{1}$=100$\,$GHz. We set the ratio of relativistic proton to electron energy k=1, as well as 1 for the filling factor, $\eta$. We measured an average redshift near F1 of z=0.2329. For  \begin{math}\mathnormal{a}\end{math}$\sim$1, P$_{min}=$2$\times$10$^{-13}$dyn/cm$^{2}$. We do the calculations assuming no relativistic boosting or deboosting, as well as including a factor to account for the effects of deboosting as in \citet{urr84}.

\begin{equation}
\mathrm{L_{boo}}=\mathrm{\delta^{2+\mathnormal{a}}L}, 
\end{equation}

where $\delta$=[$\gamma$(1-$\beta$cos($\theta$))]$^{-1}$ and $\beta$ is defined as above, $\gamma$ is the Lorentz factor, L is the luminosity at 1.4$\,$GHz and $\theta$ is the angle between the line-of-sight and the velocity vector (see Figure~\ref{bragn}).  Relativistic deboosting decreases the observed internal jet pressure so the corrected value of (8$\pm$3.2)$\times$10$^{-13}$dyn/cm$^{2}$ is more realistic. The errors are propagated from the measurement errors on the flux density and $\beta$.

The source, marked with a star in Figure~\ref{posvel}a, is traveling mainly orthogonal to the direction of the filament. The filament direction is labeled F in Figure~\ref{bragn}. To obtain its velocity relative to the filament, we considered the distribution of velocities relative to the median redshift of the 29 galaxies in the redshift limits of the structure (z=0.225-0.24) and within 1.5Mpc of F1 (Fadda et al 2008; Edwards et al 2010a). This produces a $v_{l.o.s.}=$ (373 $\pm$ 30)$\,$km/s, highlighted in Figure~\ref{posvel}b with an arrow. We deproject by dividing by cos($\theta$), where $\theta$ is 90-($\Phi$+$\alpha$/2), obtaining a velocity of 980 $\pm$ 80$\,$km/s.

\begin{figure}
   \center
  \epsscale{1.0}
     \plotone{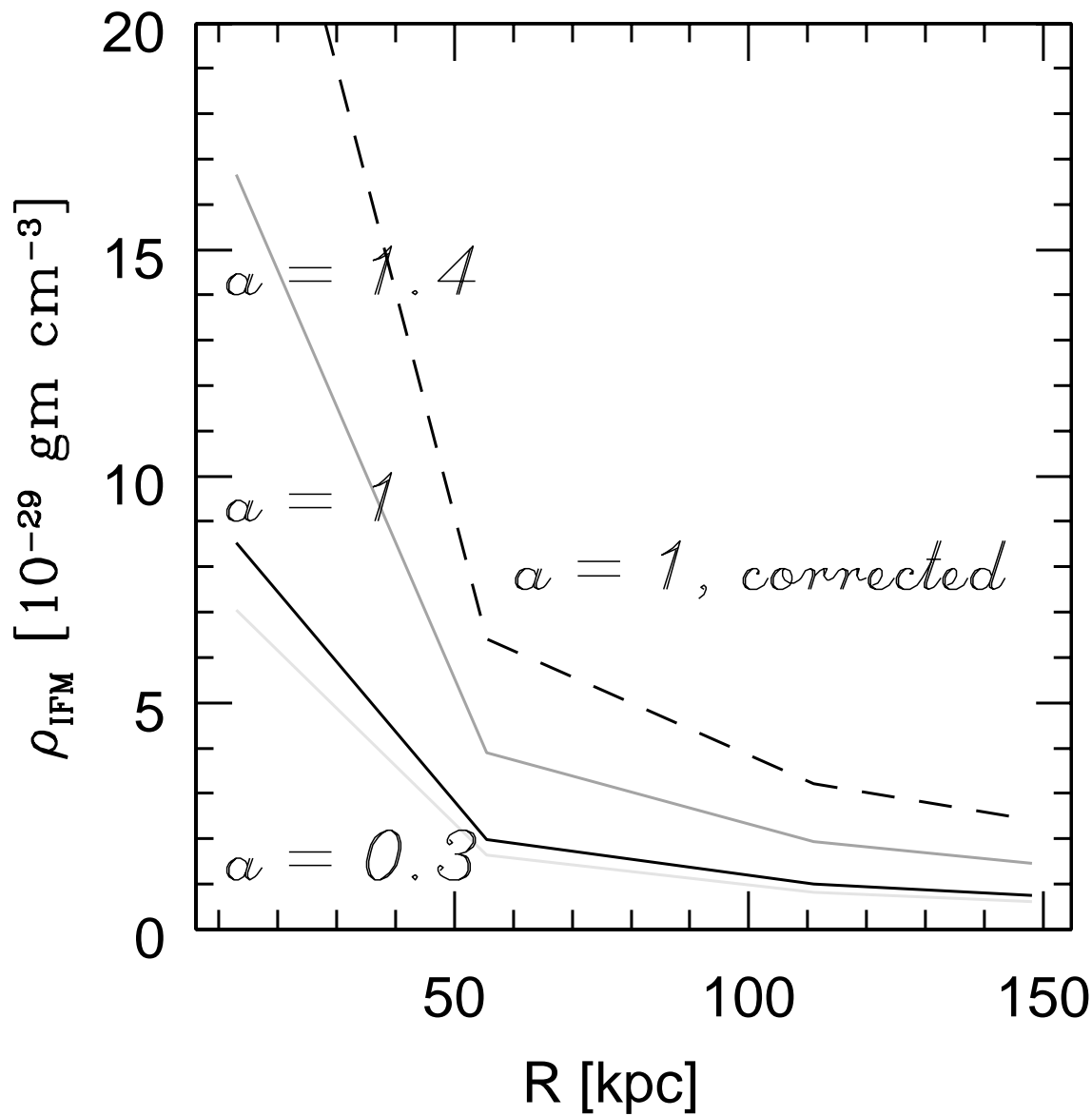}
    \caption[]{{\bf IFM density as a function of curvature radius} The IFM density for several values of spectral index, \begin{math}\mathnormal{a}\end{math} are shown as a function of the curvature radius. The dashed line shows the value corrected for relativistic deboosting. The DLRSs in the groups of \citet{fre08} have R$\sim$60$\,$kpc. Beyond R$\sim$100$\,$kpc, the $\rho_{IFM}$ begins to stabilize.
\label{rhoifm}}
\end{figure}

Figure~\ref{rhoifm} shows $\rho_{IFM}$ for select values of the spectral index, \begin{math}\mathnormal{a}\end{math} as a function of R, which begins to stabilize after R$\sim$100$\,$kpc.  The most reasonable choice of \begin{math}\mathnormal{a}\end{math} is probably around 1, and the dashed line includes a correction for relativistic deboosting. This gives $\rho_{IFM}$ between (1-20)$\times$10$^{-29}$$\,$gm/cm$^{3}$. For R = 60$\,$kpc, $\rho_{IFM}$ = 6.4$^{+14.3}_{-5.1}$$\times$10$^{-29}$$\,$gm/cm$^{3}$.

\subsection{Galaxy galaxy mergers along the filament}
We argue that there has been no significant merger activity that would affect the morphology of the jets.  First, the host galaxy of our bent DLRS shows nearly symmetric isophots, and red colors with (u$^{\prime}$-r$^{\prime}$)=2.60, with no clear indications of recent starburst activity in the optical spectrum. Second, although there is a projected close pair only 24$\,$kpc away (SDSS photometric redshift of 0.029), the merger timescale at this distance is significantly longer ($\sim$1$\,$Gyr) than the particle lifetimes in the jet ($\sim$2$\,$Myr).

\section{Discussion}

\citet{tec10} calculate the ram pressure in rich clusters at z$<$0.5 to be 2$\times$10$^{-11}$ dyn/cm$^{2}$ and an order of magnitude less for poorer clusters of M$_{vir}$$\sim$10$^{14}$$\,$M$_{\odot}$. Abell~1763 is a massive galaxy cluster with central pressures on the order of 2$\times$10$^{-10}$ dyn/cm$^{2}$ and of 10$^{-11}$ dyn/cm$^{2}$ at 1$\,$Mpc from the cluster center \citep{cav09}. At the distance of F1, we infer the minimum jet pressure to be between (8.0$\pm$3.2)$\times$10$^{-13}$ dyn/cm$^{2}$ for \begin{math}\mathnormal{a}\end{math}=1, almost 2 orders of magnitude below this. The measured electron density for this cluster at 1$\,$Mpc from the center is 5$\times$10$^{-4}$ cm$^{-3}$. For reasonable assumptions of the spectral index and for R=60$\,$kpc we have estimated $\rho_{IFM}$ between 6.4$^{+14.3}_{-5.1}$$\times$10$^{-29}$ gm/cm$^{3}$ which gives gas number densities of 6$^{+14}_{-5}$$\times$10$^{-5}$ cm$^{-3}$ assuming a mean molecular weight of 0.6. 

\nocite{car06}

Direct observations of the X-ray emission in the filament that joins Abell~222 and 223 from \citet{wer08} measure the IFM to be 3$\times$10$^{-5}$cm$^{-3}$, which agrees with our value to within errors. Although it is possible we are measuring the intra-group medium of an infalling group within the filament, the fact that our values are near to those found in the Abell~222 filament suggest we are indeed measuring the IFM. Our results are also of the same order of magnitude as those of \citet{fan10}. These authors have calculated the Warm Hot Intergalactic Medium (WHIM) in the Sculptor wall to be 30 times the mean density of the Universe, measured by \citet{rio91} to be between (2-5)$\times$10$^{-31}$ gm/cm$^{3}$. Our results are in agreement with the simulations of \citet{dav01} who predict the density of the WHIM to be 4$\times$10$^{-6}$ to 1$\times$10$^{-4}$ cm$^{-3}$, and consistent with the simulations of \citet{dol06} which predict filament densities $\sim$ 10-100 times the mean density (at distances between 2-10$\,$Mpc of a massive cluster).

\section{Conclusions}

We have found a DLRS 3.4$\,$Mpc from the core of Abell~1763 along the galaxy filament, towards Abell~1770, but well beyond the virial radius of either cluster. Geometrical arguments propel us to surmise an intrinsic bend of $\alpha =$25$^{o}$ between the jets of F1, seen at $\Phi =$10$^{o}$ with respect to the line-of-sight. We have ruled out any significant affect on the radio morphology from either an interaction with the host, or with another galaxy. Bent DLRSs have been found to reside in groups and postulated to exist in filaments \citep{fre08}. Because of our extremely wide field, deep, multiwavelength study of this super structure, {\it we know a priori that the galaxy is ultimately part of the cluster feeding filament}. This is the first time such an object has been found so far from the cluster center and identified specifically within a filament. This unique finding has led us to be able to fix observationally motivated constraints on the IFM, which we deduce to be between (1-20)$\times$10$^{-29}$ gm/cm$^{3}$. This is consistent with theoretical estimates for filament densities, as well as recent X-ray observations of cluster scale filaments. These observations serve to motivate future measures of the IFM by justifying searches for bent DLRGs in known filaments.

\citet{bla01}, starting with a sample of bent DLRSs, find that 80\% reside in clusters. We suggest such sources with distances of several Mpc from a rich galaxy cluster center, as markers for rigorous supercluster scale filament searches. We conclude that many of the bent DLRS found in groups may actually be part of larger scale structures, such as filaments. 

\acknowledgments

We thank the scientific staff at the NRAO, in particular G. Van Moorsel for help with observation planning and F. Owen for help with data reduction techniques. We thank L. Storrie-Lombardi and P. Appleton for fruitful discussions.  Support for this work was provided by NASA through an award issued by JPL/Caltech. 

This work is based in part on original observations using the Very Large Array operated by the NRAO. The National Radio Astronomy Observatory is a facility of the National Science Foundation operated under cooperative agreement by Associated Universities, Inc.
{\it Facilities:} VLA, WIYN (Hydra)

\end{document}